\documentclass[runningheads,a4paper]{llncs}

\usepackage{amssymb}
\usepackage{graphicx}
\usepackage{url}
\usepackage[tight,footnotesize]{subfigure}
\urldef{\mailsa}\path|vesna.vuksanovic@tu-berlin.de, phoevel@physik.tu-berlin.de| 
\newcommand{\keywords}[1]{\par\addvspace\baselineskip
\noindent\keywordname\enspace\ignorespaces#1}

\begin{document}

\mainmatter  

\title{Large-scale neural network model for functional networks of the human cortex
\footnote{To appear as: Vesna Vuksanovic and Philipp H\"ovel:
\textit{Modelling functional connectivity} in A. Pelster and G. Wunner (Editors): \textit{Proceedings of the International Symposium Selforganization in Complex Systems: The Past,
Present, and Future of Synergetics}; Hanse Institute of Advanced Studies, Delmenhorst, November 13 -- 16, 2012; Springer Series Understanding Complex Systems, Springer, in preparation.}}

\titlerunning{Modelling functional connectivity}

%
%
\author{Vesna Vuksanovi\'{c}\inst{1,2}%
\and Philipp H{\"o}vel\inst{1,2,3}}
%

\institute{Institut f{\"u}r Theoretische Physik, Technische Universit\"at Berlin\\
Hardenbergstra\ss{}e 36, 10623 Berlin, Germany,
\and
Bernstein Center for Computational Neuroscience, Humboldt-Universit{\"a}t zu Berlin\\
Philippstra{\ss}e 13, 10115 Berlin, Germany
\and
Center for Complex Network Research, Northeastern University, 110 Forsyth Street, Boston, USA\\
\mailsa\\}

%
%

\toctitle{Modelling functional connectivity}
\maketitle

\begin{abstract}
We investigate the influence of indirect connections, interregional distance and collective effects on the large-scale functional networks of the human cortex. We study topologies of empirically derived resting state networks (RSNs), extracted from fMRI data, and model dynamics on the obtained networks. The RSNs are calculated from mean time-series of blood-oxygen-level-dependent (BOLD) activity of distinct cortical regions via Pearson correlation coefficients. We compare func\-tional-connectivity networks of simulated BOLD activity as a function of coupling strength and correlation threshold. Neural network dynamics underpinning the BOLD signal fluctuations are modelled as excitable FitzHugh-Nagumo oscillators subject to uncorrelated white Gaussian noise and time-delayed interactions to account for the finite speed of the signal propagation along the axons.  We discuss the functional connectivity of simulated BOLD activity in dependence on the signal speed and correlation threshold and compare it to the 
empirical data.
\keywords{functional connectivity, resting state networks, time-delays}
\end{abstract}
\section{Introduction}
Despite important progress over past few years, little is known about brain functional connectivity (FC) at rest, i.e. under no stimulation and in the absence of any overt-directed behaviour. 
In the studies on goal-directed mental activity, spontaneous brain activity has been considered as random enough to be averaged out across many trials \cite{FOX07a}. 
However, well organized spatio-temporal low-frequency fluctuations ($<$ 0.1 Hz) have been observed in blood-oxygen-level-dependent (BOLD) fMRI signals of a mammalian brain in the absence of any stimulation or task related behaviour \cite{BIS95,DAM06,VIN07a}. 
These well organized patterns of activity, suggest existence of underlying dynamics that governs intrinsic brain processes. 

Existing large-scale models of the intrinsic brain dynamics focus on the relationship between functional and anatomical connectivity. They explore the range of conditions under which functional networks may emerge from known anatomical connections. In particular, roles of multiple time-scales in FC networks formation \cite{HON07,RUB09}, time-delays in the signal propagation between the network nodes and system noise \cite{GHO08,GHO08a}, local network oscillations \cite{DEC09,CAB11} and structural disconnection \cite{CAB12} in the underlying dynamics have been investigated. Although RSNs reflect anatomical connectivity between brain areas comprising the networks in focus, FC cannot be understood in those terms alone \cite{BUL09,DEC12}.

Here, we combine experimental and modelling approaches to investigate dynamics underlying correlated behaviour of distant cortical regions. We choose to model the local node dynamics by excitable FitzHugh-Nagumo (FHN) neurons \cite{FIT61,NAG62} and use the resulting time-series to infer the neuronal activity at the network levels and subsequently low-frequency oscillations in the BOLD data \cite{FRI00}. We focus on identifying
how global coupling strength and different network topologies affect connectivity patterns in simulated BOLD signals. 

The rest of this paper is organized as follows: In Sec.~\ref{sec:model} we provide a detailed description of the methods used to generate the networks. In addition we introduce the model equations. This sets the stage for numerical simulations that are presented in Sec.~\ref{sec:results}. Finally we conclude with a discussion in Sec.~\ref{sec:discussion}.
\section{Methods and Models}
\label{sec:model}

\textbf{Subjects.} We use resting state blood-oxygen-level-dependent (BOLD) fMRI signals downloaded from the \textit{1000 Functional Connectome Project} website (\url{http://www.nitric.org/}). 26 functional and anatomical scans from the Berlin study are considered in the analysis. Demographic data for the subjects participated in the study along with the technical details of the signal acquisition can be found at the \textit{Functional Connectome} website.

\textbf{Data Preprocessing Steps in FSL.} Image preprocessing is carried out using FSL (\url{http://www.fmrib.ox.ac.uk}) \cite{JEN12}.
The preprocessing consists of the following steps: (a) temporal high-pass filtering (using Gaussian-weighted least-squares straight line fitting with sigma equals 100 s); (b) temporal low-pass filtering (using Gaussian filter with HWHM = 2.8 s); (c) slice-timing correction for interleaved acquisition (using Fourier-space time-series phase-shifting); (d) motion correction (using a six parameter affine transformation implemented in \textsc{MCFLIRT}); (e) spatial filtering (using Gaussian kernel of FWHM = 6 mm); (f) non-brain removal (BET brain extraction is applied to create a brain mask from the first volume in the fMRI data): (g) normalization to standard brain image (using 12 DOF linear affine transformation implemented in \textsc{FLIRT} each subject scan was transformed in MNI space - voxel size $2\times2\times2$ mm).

\textbf{Functional Connectivity.} For the connectivity analysis we extract BOLD time-series from $N = 64$ cortical regions. These regions are adapted from studies of functional segmentation of the human cortex \cite{KIV09,AND11}. They consist of 30 pairs of the inter-hemispheric homologues and 4 additional areas chosen along the midline. The full list of the cortical areas along with their anatomical (MNI) coordinates is provided in the work by Anderson et al. in Table 1 in the Supplementary material of Ref.~\cite{AND11}. We define the areas of interest as 5 mm spheres centered at the corresponding anatomical coordinates. For the given image resolution of $2\times2\times2$ mm voxel size (measured at $64\times64\times33$ brain sites in total), there are 81 voxel time-series within each of the spherical areas. To obtain correlation matrix, describing connections between the regions, we calculate the linear correlation coefficient between any pair of the mean signal intensity of each of the 64
regions as:
\begin{equation} 
  r (x_1,x_2) = \frac{\langle V(x_1,t)V(x_2,t)\rangle - \langle V(x_1,t)\rangle \langle V(x_2,t)\rangle}{\sigma\left(V(x_1,t)\right) \sigma\left(V(x_2,t)\right)},
\end{equation}
where $V(x,t)$ represents activity of the region $x$ at time $t$ (averaged across 81 voxels), $\sigma$ is the standard deviation, 
and $\langle \cdot \rangle$ denotes temporal averages. The
$N \times N$ correlations matrices are averaged across the ensemble of 26 subjects. The obtained correlation matrix $\{f_{ij}\}$, $i,j=1,\dots,N$, is used to create FC maps between the cortical regions of interest. By definition, FC between two brain regions exists if their temporal correlation exceeds some predetermined value $r$, regardless of their anatomical connectivity. 

\textit{Thresholding:}
We obtain FC network topologies by thresholding the empirically derived FC matrix at different threshold values $r$. If $f_{ij} \geq r$, the corresponding element of the adjacency matrix is set to 1; otherwise it is set to 0. We simulate neural and BOLD activity only for $r \geq 0.26$, i.e. values for which connection density or topology cost in the brain networks is less than 0.5 \cite{BUL11a}.

\textbf{Simulation of the Network Dynamics.}
To infer the BOLD signal in dependence on the network properties, we first simulate the underlying neural activity. We consider the neural dynamics on the network of $N$ nodes or cortical regions, where 
local dynamics of the each node is represented by the homogeneous FHN neurons. 

\begin{figure}
\centering
\includegraphics[width=0.6\textwidth]{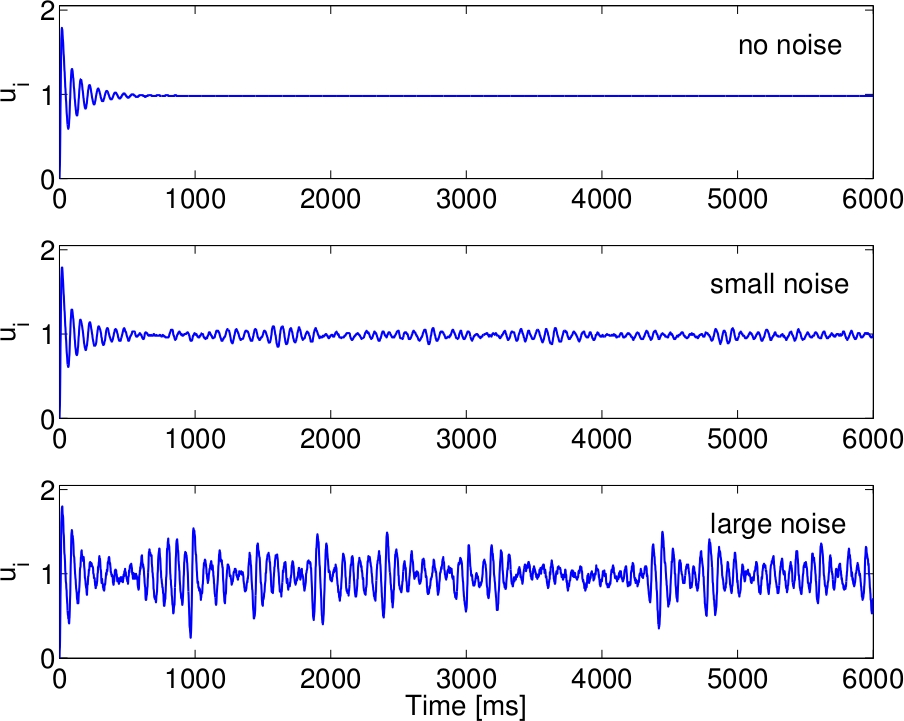}
\caption{Time evolution of the activator variable of the FitzHugh-Nagumo neural model for an isolated node under noise with different intensity.}
\label{fig:Fig1}
\end{figure}

\begin{figure}
\centering
\begin{minipage}{.48\textwidth}
  \centering
  \includegraphics[width=.8\linewidth]{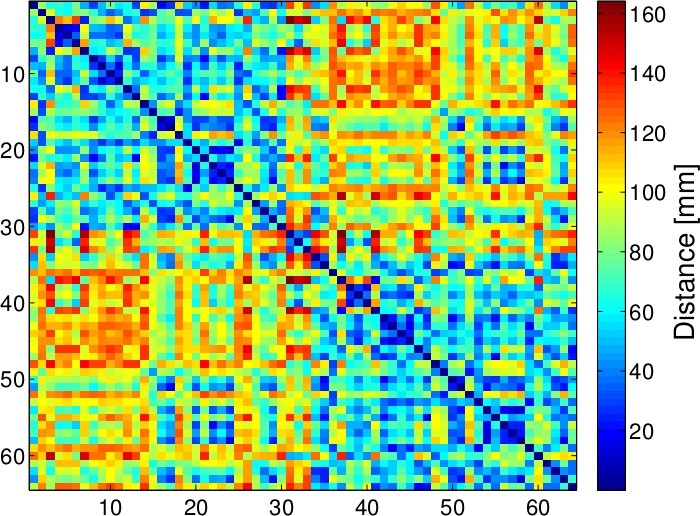}
  \caption{Distance matrix $\{d_{ij}\}$, $i,j=1,\dots,N$, between cortical regions in colour code. $d_{ij}$ are calculated as Euclidean distance between centers of spherical regions in anatomical space. Matrix is ordered in a way that corresponding contra-lateral regions are symmetrically arranged with respect to the matrix centre (right hemisphere regions, nodes 1 - 30; left hemisphere regions, nodes 35 - 64). 4 regions selected along the midline are placed in the middle of the matrix (nodes 31 - 34).}
  \label{Fig2}
\end{minipage}%
\hspace{.02\textwidth}
\begin{minipage}{.48\textwidth}
  \centering
  \includegraphics[width=.95\linewidth]{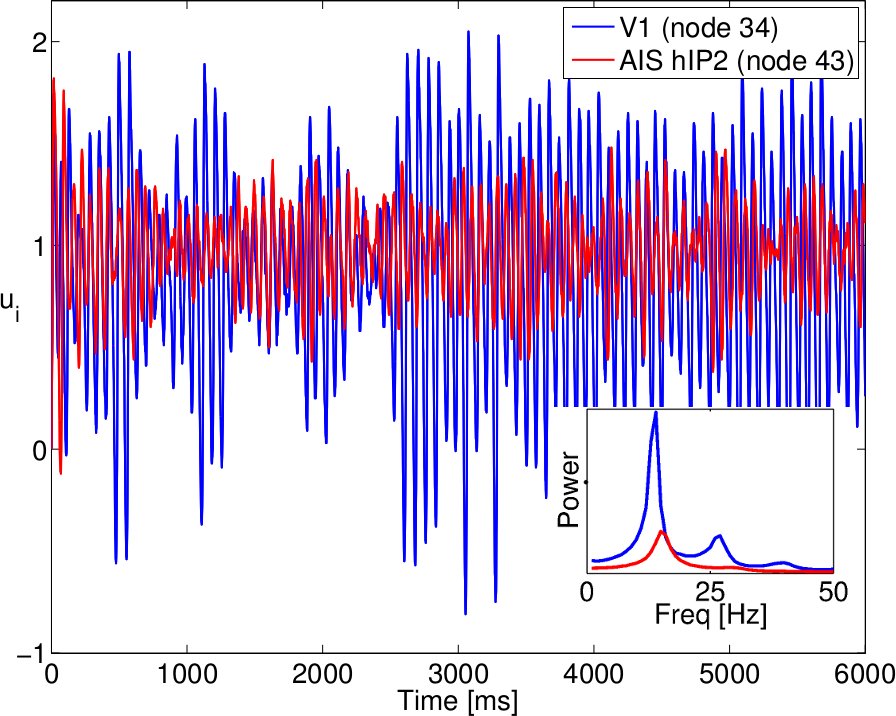}
  \caption{Time-series of neural network nodes dynamics for one highly and one sparsely connected nodes in visual cortex (V1, blue) and anterior intraparaetal sulcus (AIS hIP2, red), respectively. Network dynamics are modelled with time-delays, large noise ($D = 0.05$) and weak coupling ($c = 0.016$). Parameters: $\alpha = 0.85$, $\beta = 0.2$, $\gamma = 1.0$ and $\tau = 1.25$. The inset shows the corresponding power spectra.\vspace{1cm}}
  \label{Fig3}
\end{minipage}
\end{figure}

\textit{Simulation of the Neural Activity:}
Simulations of the neural network dynamics take into account time-delays due to finite speed of signal propagation between the nodes as well as the presence of the system noise:
\begin{equation} 
   \dot{u}_i = g(u_i,v_i) - c {\sum_{j=1}^N f_{ij}u_j(t - \Delta t_{ij})} + n_u\\
\end{equation}
\begin{equation}
   \dot{v}_i = h(u_i,v_i) + n_v,
\end{equation}
where $c$ denotes global coupling parameter ($c > 0$), $u_i$ and $v_i$ are the activator and inhibitor variables of the $i^{th}$ neural population, $f_{ij}$ are the elements of the FC matrix and $\Delta t_{ij}$ denote time-delays. $n_u$ and $n_v$ are two independent additive white Gaussian noise terms with zero mean, unity variance and noise strength $D$. The functions $g$ and $h$ describe the local neural activity according to the FHN model, where we use the notation of Refs.~\cite{GHO08,GHO08a}:

\begin{equation} 
  \dot{u} = g(u,v) = \tau \left(v + \gamma u - \frac{u^3}{3}\right)\\
\end{equation}
\begin{equation}
  \dot{v} = h(u,v) = -\frac{1}{\tau}\left(u - \alpha + bv -I\right),
\end{equation}
where $I$ is magnitude of an external stimulus, which is assumed to be 0 in our model of the resting state dynamics \cite{GHO08}. The system parameters are chosen as $\alpha = 0.85$, $\beta = 0.2$, $\gamma = 1.0$ and $\tau = 1.25$ to render solutions with damped oscillatory behaviour of a node dynamics, i.e. at the onset of instability, in the absence of the connectivity in the network \cite{GHO08,GHO08a}. Representative time-series of the dynamics of an isolated node are shown in Fig.~\ref{fig:Fig1} for different noise strengths $D=0$ (no noise), $D=0.01$ (small noise) and $D=0.05$ (large noise). 
We solve the system of coupled differential equations with time-delays and additive noise using the \textsc{Python}-module \textsc{pydelay} \cite{FLU09a}. The algorithm is based on the Bogacki-Sampine method \cite{BOG89,SHA01a}, which is also implemented in \textsc{Matlab's dde23}. We calculate time-delays for a physiologically realistic value of the propagation velocity ($v$ = 7 m/s) via $\Delta t_{ij} = d_{ij}/v$, where $d_{ij}$ are approximated by the Euclidean distances between the centers of the spherical regions, i.e. network nodes $i$ and $j$.
A colour-coded representation of the $d_{ij}$ values is shown in Fig.~\ref{Fig2}. We simulate 7.5 minutes of the real time using a time step of 1 ms.  When time-delays, system noise and global coupling are taken into account, the simulated dynamics of the neural activity exhibit a behaviour that are exemplary shown in Fig.~\ref{Fig3}.

\textit{Simulation of the BOLD Activity:}
To infer the BOLD activity from the simulated neural activity we assume the Baloon-Windkessel hemodynamic model \cite{FRI00}. Simulated BOLD signals are band-passed in the frequency interval (0.01 - 0.15) Hz to match low-frequency oscillations present in the experimental BOLD data and finally down-sampled to 2.3 s to match the temporal resolution of the MRI scanner.  
\section{Results}
\label{sec:results}

\begin{figure}
  \centering
  \begin{subfigure}
  \centering
   \includegraphics[width=.3\linewidth]{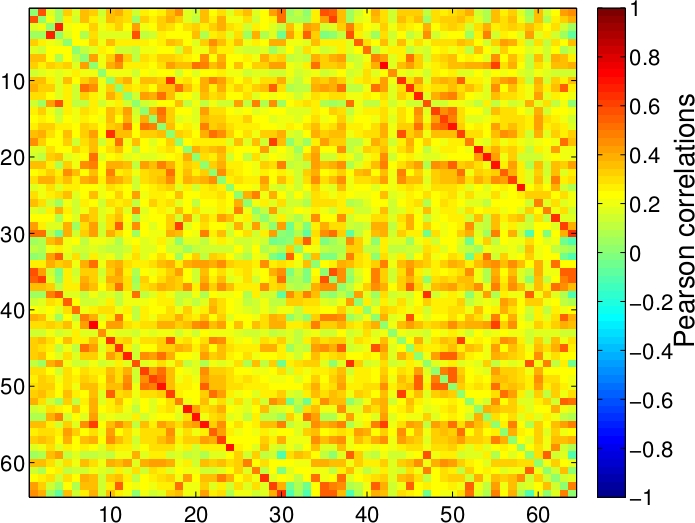}
   \end{subfigure}%
  \begin{subfigure}
  \centering
  \includegraphics[width=.65\linewidth]{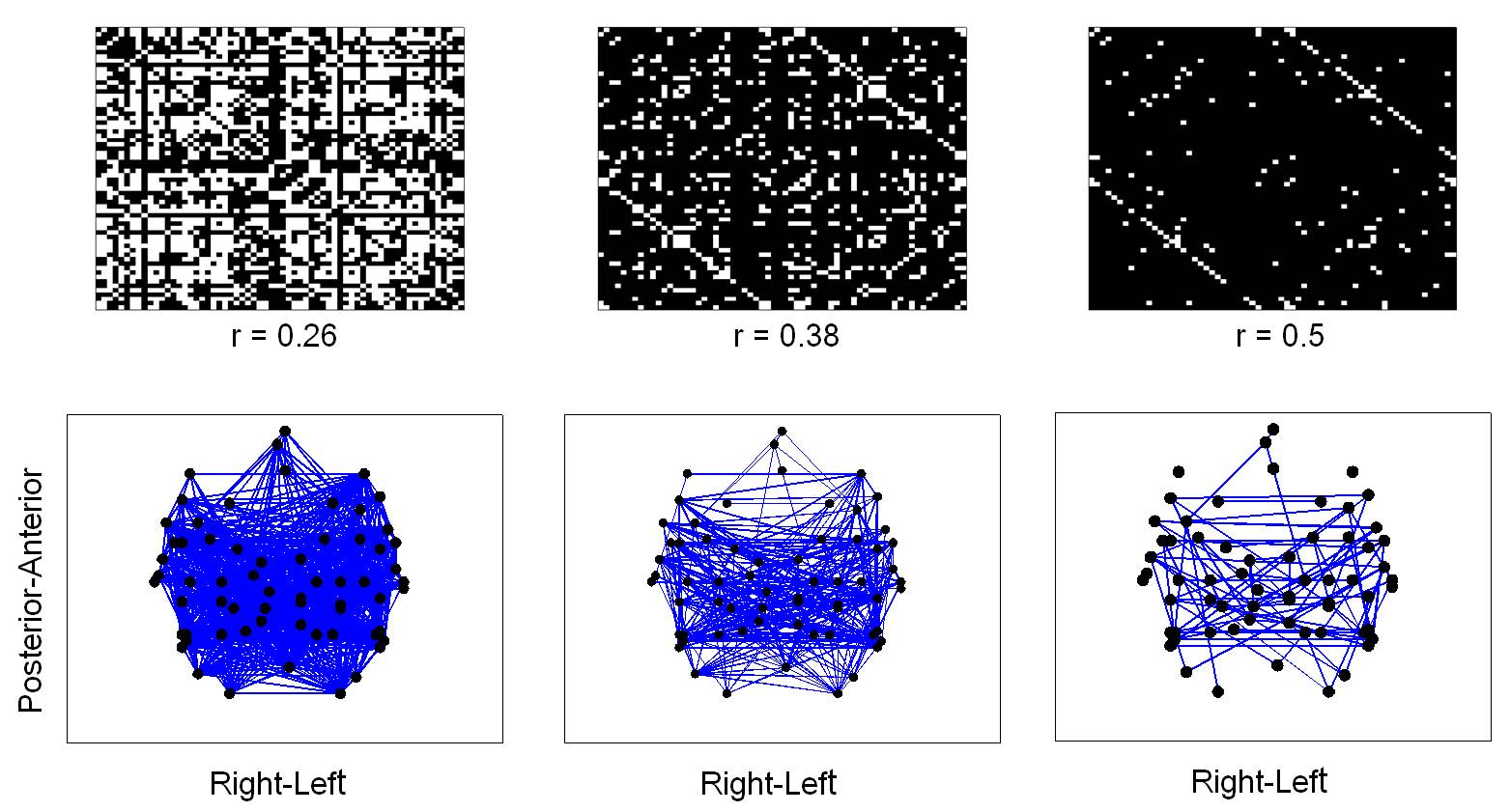}
\end{subfigure}
\caption{{\bf(Left)} Functional connectivity matrix constructed by calculating Pearson correlations on all pairwise combinations of the BOLD data from 64 cortical regions. The matrix is ordered as in the Fig.~\ref{Fig2}. The anti-diagonals reveal existing high correlations between contra-lateral regions. {\bf(Right)} Top panels: Binarized functional-connectivity matrix at thresholds $r = 0.26$, $0.38$, and $0.5$. Each element is either black (if there is no significant connections between the regions) or white (if there is). Bottom Panel: Visualization of thresholded matrices in anatomical space by locating each region according to its $x$ and $y$ coordinates and drawing a link between significantly connected regions.}
\label{fig:Fig4}
\end{figure}

\textbf{Empirical Functional Connectivity.} FC matrices obtained from fMRI data and its binarized versions for different thresholds are shown in Fig.~\ref{fig:Fig4} (left) and the top panels Fig.~\ref{fig:Fig4} (right), respectively. In the latter figure, the threshold values are chosen as $r = 0.26$, $0.38$ and $0.5$. $r = 0.26$ represents mean value of all correlations in the empiricaly derived FC and also the value at wich connection density ($\kappa$) of the network drops below 0.5, ($\kappa = 0.48$). Values $r = 0.38$ and $0.5$ are one and two standard deviations away from the mean, respectivelly.
Horizontal-plane view illustrate the anatomical maps of the corresponding networks structure in the bottom panels of the Fig.~\ref{fig:Fig4} (right). In this representation each cortical region is treated as a network node with the connections or links to the other nodes in network if the matching correlation exceeds given threshold. 
FC networks show a pronounced left to right symmetry for different values of the correlation thresholds. However, distributions of the areas with most connections depend on the threshold applied. The areas showing high numbers of links for the lower threshold network differ from the areas that appear as highly connected at higher thresholds. We expect that nodes with increased connectivity play an important role in the system's dynamics.

\begin{figure}
\centering
\includegraphics[height=6.8cm]{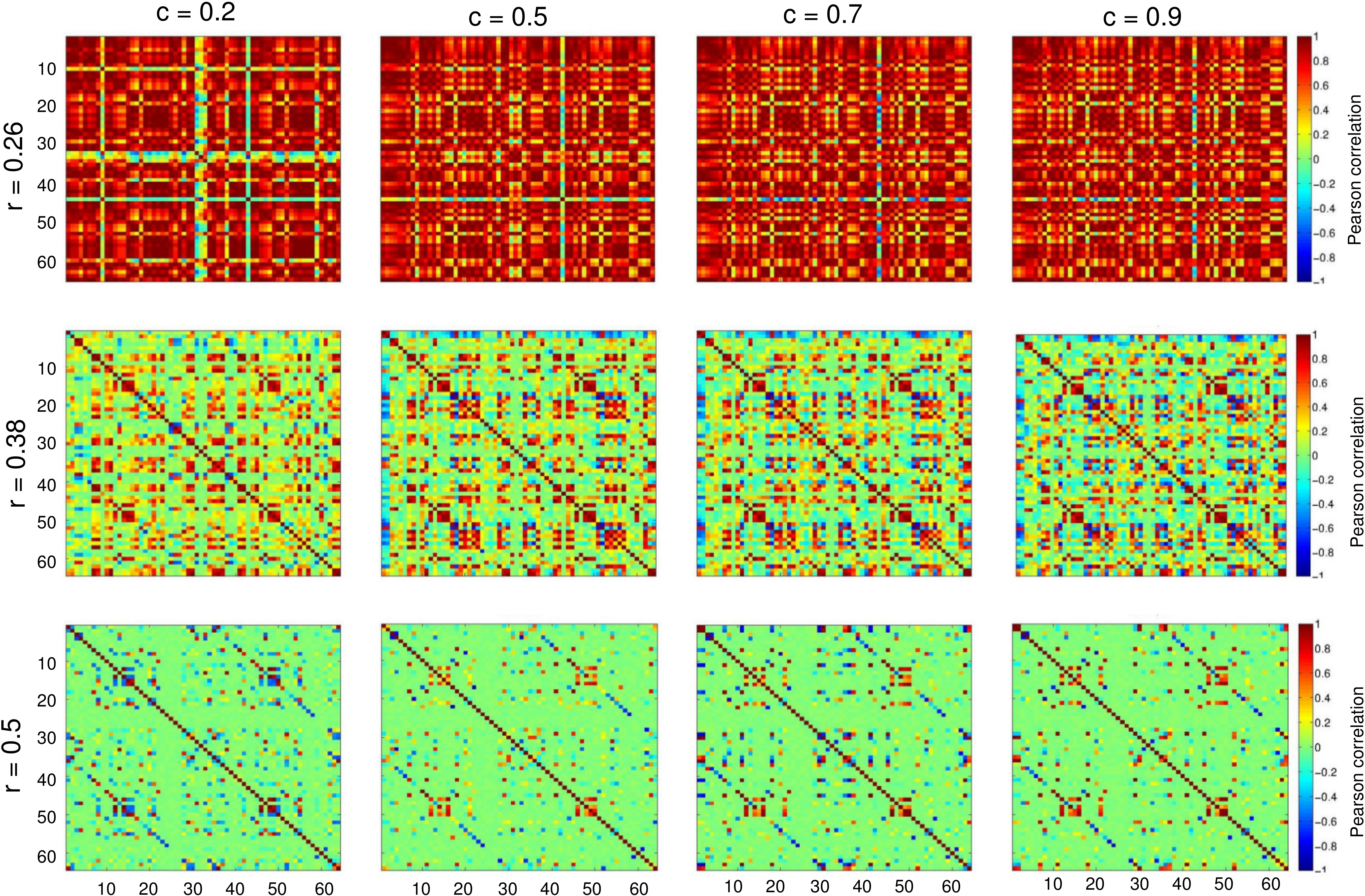}
\caption{Functional connectivity (FC) of simulated BOLD activity for different correlation thresholds $r$ applied to the empirically derived FC matrix and different coupling strengths $c$. Other parameters as in Fig.~\ref{Fig3}.}
\label{Fig5}
\end{figure}

\textbf{Simulated Functional Connectivity.} FC of the simulated BOLD activity are shown in Fig.~\ref{Fig5}. The data are obtained varying correlation threshold $r$, i.e. network topology and global coupling strength $c$ while keeping other model parameters at constant values. Correlations in the brain FC networks may not necessarily originate from direct connections. Therefore, we vary the network topologies by changing the correlation threshold $r$ in the binary filter applied to the empirically derived FC network. We use the corresponding FC matrices to study the dynamics that result in the  appearance of these correlations. Furthermore we uniformly scale all connections strengths $c$ between the network nodes. For low correlation thresholds and weak coupling strengths ($c \leq 0.05$) the simulated BOLD signals are weakly correlated as the underlying neuronal activity does not show correlated behaviour either (data not shown). Increasing the coupling strength, positive correlations among the BOLD signals 
emerge in the simulated data and 
corresponding FC networks exhibit patterns 
of correlated activity similar to the empirically derived FC.
\section{Discussion/Conclusion}
\label{sec:discussion}
In this study we presented data obtained by combining both experimental and modelling approaches to explore dynamics underlying correlated behaviour of distant cortical regions. We showed that increasing global coupling strength between the network nodes introduces correlations into simulated BOLD activity. The spatial patterns of the correlated activity resembled those observed in corresponding experimental data. However, to get tangible measures of similarity between experimentally and theoretically derived FC networks our next step would be to apply graph-theoretical analysis to the obtained data.

Our results are consistent with recent findings showing that stronger structural couplings generates more globally connected and globally integrated BOLD signals \cite{CAB12}. Since functional connectivity implies possible role of indirect anatomical connections further studies, showing how our findings relate to known anatomical connections between cortical regions could lead to the better understanding of FC networks formation. 
\subsubsection*{Acknowledgments.}
This work was supported by BMBF (grant no. 01Q1001B) in the framework of BCCN Berlin (Project B7). We thank John-Dylan Haynes and his group for helpful discussions concerning the data processing.

\end{document}